\documentclass[options]{JHEP3} 
\usepackage{indentfirst}

\usepackage{epsfig,multicol}

\newcommand\fverb{\setbox\pippobox=\hbox\bgroup\verb}
\newcommand\fverbdo{\egroup\medskip\noindent%
			\fbox{\unhbox\pippobox}\ }
\newcommand\fverbit{\egroup\item[\fbox{\unhbox\pippobox}]}
\newbox\pippobox

\title{Quantum effective potential for $U\left(1\right)$ fields on ${\bf S}^{2}_{L}\times {\bf S}^{2}_{L}$}

\author{P.Castro-Villarreal$^{+,*}$ , R.Delgadillo-Blando$^{+,*}$ , Badis Ydri$^{+,**}$\\
$^{+}$ School of Theoretical Physics, Dublin Institute for Advanced Studies, Dublin, Ireland.\\
$^{*}$ Dept. de F\'{\i}sica, Centro de  Investigaciones  de Estudios Avanzados del IPN,\\ ~~Apdo. Postal 14-740, 07000, M\'exico D.F., M\'exico.\\
$^{**}$Department of Mathematical Physics, NUI Maynooth, Ireland.}

\received{} 		
\revised{}
\accepted{}		

\preprint{}	

\abstract{We compute the one-loop effective potential for noncommutative $U(1)$ gauge fields on ${\bf S}^2_L{\times}{\bf S}^2_L$. We show the existence of a novel phase transition in the model from the  $4-$dimensional space ${\bf S}^2_L{\times}{\bf S}^2_L$ to a matrix phase where the spheres collapse under the effect of quantum fluctuations.  It is also shown that the transition to the matrix phase occurs at infinite value of the gauge coupling constant when the mass of the two normal components of the gauge field on ${\bf S}^2_L{\times}{\bf S}^2_L$ is sent to infinity.}

\keywords{Fuzzy sphere, NC gauge theory, matrix model, effective action, phase structure}

\dedicated{}

\begin{document}

\section{Introduction}

Fuzzy approximations of spacetime (like lattice regularizations) are designed for the study of gauge theories 
in the nonperturbative regime using Monte-Carlo simulations. They consist in replacing continuous manifolds
by matrix algebras. The resulting field theory will thus only have  a finite number of 
degrees of freedom and hence it is regularized. The claim is that this method has the advantage 
-in contrast with lattice- of preserving all continuous symmetries of the original action at the classical level \cite{madore,peter,thesis}.

Field theory on the fuzzy sphere is the most studied example in the
literature. In perturbation theory it is shown that scalar field theories on ${\bf S}^2_L$ suffer from the UV-IR mixing problem \cite{scalar}.  Moreover it is shown that there exists
new nonperturbative phenomena which are missing in the commutative
theory. For example a novel phase has been
discovered in scalar field theories on ${\bf S}^2_L$ (the so-called
non-uniform phase or matrix phase) which has no commutative analogue
\cite{xavier}. This new phase was also observed in three dimensions
\cite{bieten}. Generalization to $4-$dimensional fuzzy spaces and their scalar field theories were undertaken in \cite{scalar4d}.

The quantum properties of the gauge field on the fuzzy sphere have been studied in \cite{PRY,nishimura,steinacker}. In \cite{PRY} the
effective action was computed to one loop for $U(1)$ gauge fields. It
was shown that the model contains a gauge invariant UV-IR mixing in
the limit $L\to \infty$, i.e the effective action does not
go over to the commutative action in the continuum limit. Furthermore a first order  phase transition  was observed at one-loop from the fuzzy sphere phase to a matrix phase where the sphere collapses. This  transition was previously detected in Monte Carlo simulation of the model reported in \cite{nishimura}.  In some sense the one-loop result for the $U(1)$ model is exact. 

It was also shown in \cite{PRY} that the UV-IR mixing and the matrix phase both disappear in the limit where we send the mass of the normal scalar component of the gauge field on ${\bf S}^2_L$ to infinity. This means in particular that the nonperturbative ${\bf S}^2_L$-to-matrix phase transition is a reflection of the UV-IR mixing seen in perturbation theory and that this latter finds its origin in the coupling of the normal scalar field to the two dimensional gauge field. The differential calculus on the fuzzy sphere is intrinsically $3-$dimensional and as a consequence there is no a gauge-covariant splitting of the $3-$dimensional fuzzy gauge field into its normal and tangent components on ${\bf S}^2_L$; hence the action will necessarily involve the interaction of the two fields. This result (among many others) was confirmed recently in our Monte Carlo simulation of the model where  we have also found a novel third-order one-plaquette-like phase transition which the model undergoes and which we can also trace to the coupling of the normal scalar field. The full phase diagram of the model will be reported elsewhere \cite{ydri}.

The main goal of this article is to study the phase structure of $U(1)$ gauge theories on fuzzy ${\bf S}^2_L{\times}{\bf S}^2_L$. The advantage of considering ${\bf S}^2_{L}\times {\bf S}^2_{L}$
is that one can use
all the machinery of the well known $SU(2)$ polarization tensors.
Other studies of noncommutative gauge theories  on $4-$dimensional fuzzy spaces have already appeared \cite{othergaugein4d}. 

This article is organized as follows. In section $2$ we give a brief description of the geometry of fuzzy ${\bf S}^2_L{\times}{\bf S}^2_L$. In section $3$ we introduce fuzzy gauge fields and we write down the action we will study in this article. In section $4$ we compute the effective potential. In section $5$ we show the existence of a first order ${\bf S}^2_L{\times}{\bf S}^2_L$-to-matrix phase transition in exact analogy with the two-dimensional case and we derive the critical line. Section $6$ contains the conclusion.

\section{Fuzzy ${\bf S}^{2}_{L}\times {\bf S}^{2}_{L}$}
Fuzzy ${\bf S}^{2}_{L}\times {\bf S}^{2}_{L}$ is the simpliest $4$ dimensional fuzzy space. It is a finite dimensional matrix approximation of the cartesian product of two continuous spheres. This fuzzy space is defined by a sequence of Connes  triples \cite{connes}
\begin{equation}
{\bf S}^{2}_{L}\times {\bf S}^{2}_{L}=\left\{Mat_{\left(L+1\right)^2},H_{L},\Delta_{L}\right\}.\label{2.1}
\end{equation}
$Mat_{\left(L+1\right)^2}$ is the matrix algebra of dimension $\left(L+1\right)^2$ and $\Delta_{L}$ is a suitable Laplacian acting on matrices which encodes the geometry of the space. It is defined by
\begin{equation}
\Delta_{L}\equiv\left[L_{AB},\left[L_{AB},\cdot~\right]\right]=\mathcal{L}^{2}_{AB}
\end{equation}
where $L_{AB}$, with $A,B=\overline{1,4}$, are the generators of the irreducible representation $(\frac{L}{2},\frac{L}{2})$ of $SO\left(4\right)$. The generators $L_{AB}$  (with $L_{AB}=-L_{BA}$)  satisfy the commutation relations
\begin{eqnarray}
\left[L_{A B},L_{C D}\right] &=& f_{A B C D E F}L_{E F}\nonumber\\
&{\equiv}& \delta_{B C}  L_{A D}-\delta_{B D}L_{A C}+
                                     \delta_{A D} L_{B C} -\delta_{A C} L_{B D}.\end{eqnarray}
$H_{L}$ in (\ref{2.1}) is the Hilbert space (with  inner product $<M,N>=\frac{1}{\left(L+1\right)^2}Tr\left(M^{\dagger}N\right)$) which is associated with the irreducible representation $(\frac{L}{2},\frac{L}{2})$ of $SO(4)$.

Since $SO(4)=\big[SU(2){\times}SU(2)\big]/Z_2$ we can introduce $SU(2)$ (mutually commuting) generators $L^{(1)}_a$ and $L^{(2)}_a$ by $-2L^{\left(1\right)}_{a}=\frac{1}{2}\epsilon_{abc}L_{bc}+L_{a4}$ and $-2L^{\left(2\right)}_{a}=\frac{1}{2}\epsilon_{abc}L_{bc}-L_{a4}$ with $a=1,2,3$ and $\epsilon_{abc}$ is the three dimensional Levi-Civita tensor. Then it can be easily shown that the two $SO(4)$ quadratic Casimir can be rewritten in the form (where $\epsilon_{ABCD}$ is the four dimensional Levi-Civita tensor) 
\begin{eqnarray}
L_{AB}^2&=&4\big[(L_{a}^{(1)})^2+(L_{a}^{(2)})^2\big]=2L(L+2){\equiv}8c_2\nonumber\\
{\epsilon}_{ABCD}L_{AB}L_{CD}&=&8\big[(L_a^{(1)})^2-(L_a^{(2)})^2\big]{\equiv}0.
\end{eqnarray}
Similarly the Laplacian $\mathcal{L}^{2}_{AB}$ reads in terms of the three dimensional indices as follows
\begin{equation}
\mathcal{L}^{2}_{AB}=4\left[\left(\mathcal{L}^{\left(1\right)}_{a}\right)^{2}+\left(\mathcal{L}^{\left(2\right)}_{a}\right)^{2}\right],
\label{laplacian}
\end{equation}
where {\small$\mathcal{L}^{\left(1\right)}_{a}\equiv [L^{\left(1\right)}_{a},\cdot~]$} and {\small $\mathcal{L}^{\left(2\right)}_{a}\equiv [L^{\left(2\right)}_{a},\cdot~]$}. For ${\bf S}^{2}_{L}\times {\bf S}^{2}_{L}$ the algebra $Mat_{\left(L+1\right)^2}$ is  generated by the coordinate operators 
\begin{equation}
x^{\left(1\right)}_{a}=R_{1}\frac{L^{\left(1\right)}_{a}}{\sqrt{c_2}},~~~~~ x^{\left(2\right)}_{a}=R_{2}\frac{L^{\left(2\right)}_{a}}{\sqrt{c_2}}\label{cor}
\end{equation}  
which satisfy
\begin{equation}
\sum^{3}_{a=1}\left(x^{\left(i\right)}_{a}\right)^{2}=R^{2}_{i}{\bf 1},~~~~~\left[x^{\left(i\right)}_{a},x^{\left(j\right)}_{b}\right]=\frac{i\;R_{i}}{\sqrt{c_2}}\delta_{ij}\epsilon_{abc}x^{\left(i\right)}_{c}, ~~~~i=1,2.
\end{equation}
In the limit $L\to\infty$ keeping $R_{1}$ and $R_{2}$ fixed we recover the commutative algebra of functions on ${\bf S}^{2}\times {\bf S}^{2}$.  If we also choose to scale the radii $R_{1}$ and $R_{2}$ such as for example $\theta^{2}_{1}=R^{2}_{1}/L_{1}$ and $\theta^{2}_{2}=R^{2}_{2}/L_{2}$ are kept fixed we obtain the non-commutative Moyal-Weyl space $\mathbb{R}^{2}_{\theta_{1}}\times\mathbb{R}^{2}_{\theta_{2}}$ \cite{scalar4d}.

 The algebra of matrices $Mat_{(L+1)^2}$ can be decomposed under the action of the two $SU(2)$ of $SO(4)$ as $Mat_{\left(L+1\right)}\otimes Mat_{\left(L+1\right)}$.  As a consequence a general function on $ {\bf S}^2_{L}\times {\bf S}^{2}_{L}$ can be expanded in terms of polarization tensors \cite{VKM} as follows
\begin{eqnarray}
{\phi}=\sum_{k_{1}=0}^{L}\sum_{m_{1}=-k_{1}}^{k_{1}}\sum_{k_{2}=0}^{L}\sum_{m_{2}=-k_{2}}^{k_{2}}{\phi}_{k_{1}m_{1}k_2 m_2 }\hat{Y}_{k_{1}m_{1}}\otimes\hat{Y}_{k_{2}m_{2}}.
\end{eqnarray}

\section{Fuzzy gauge fields}

$U(n)$ gauge field on ${\bf S}^2_L{\times}{\bf S}^2_L$ can be associated with a set of six hermitian matrices $D_{AB}\in Mat_{n(L+1)^2}$ ($D_{AB}=-D_{BA}$) which transform homogeneously under the action of the group, i.e
\begin{equation}
D_{AB}\to UD_{AB}U^{-1}, ~~~~U\in U\left(n(L+1)^2\right).
\end{equation}
In this paper we will be  mainly interested in  $U(1)$ theory on $S^{2}_{L}\times S^{2}_{L}$. The action is given by (with $Tr_{L}=\frac{1}{\left(L+1\right)^{2}}Tr$, $g$ is the gauge coupling constant and $m$ is the mass of the normal components of the gauge field )

\begin{eqnarray}
S& = & \frac{1}{16g^2}\left\{-\frac{1}{4}Tr_{L}[D_{AB},D_{CD}]^2+\frac{i}{3}f_{ABCDEF}Tr_{L}[D_{AB},D_{CD}]D_{EF}\right\}\nonumber\\
  &   &  \nonumber \\
  & + &  \frac{m^2}{8g^2L^2_{AB}}Tr_{L}(D^2_{AB}-L^2_{AB})^2+\frac{m^2}{32g^2L^2_{AB}}Tr_{L}(\epsilon_{ABCD}D_{AB}D_{CD})^2.
\label{model}
\end{eqnarray}
The equations of motion are given by
\begin{eqnarray}
i[D_{CD},F_{AB,CD}]+
\frac{4m^2}{\sqrt{c_2}}\{D_{AB},{\Phi}_1+{\Phi}_2\}+\frac{m^2}{\sqrt{c_2}}\{\epsilon_{ABCD}D_{CD},{\Phi}_1-{\Phi}_2\} =0.\label{21}
\end{eqnarray}
As we will see shortly $F_{AB,CD}=i\left[D_{AB},D_{CD}\right]+f_{ABCDEF}D_{EF}$ can be interpreted as the curvature of the gauge field on fuzzy ${\bf S}^{2}_{L}\times {\bf S}^{2}_{L}$ whereas ${\Phi}_1$ and ${\Phi}_2$ (defined by $D_{AB}^2-L_{AB}^2=8\sqrt{c_2}({\Phi}_1+{\Phi}_2)$ and ${\epsilon}_{ABCD}D_{AB}D_{CD}=16\sqrt{c_2}({\Phi}_1-{\Phi}_2)$) can be interpreted as  the  normal components of the gauge field on ${\bf S}^{2}_{L}\times {\bf S}^{2}_{L}$.


The most obvious non-trivial solution of the equations of motion (\ref{21})  must satisfy $F_{AB,CD}=0$, $D^{2}_{AB}=L^{2}_{AB}$ and $\epsilon_{ABCD}D_{AB}D_{CD}=0$ (or equivalently $F_{AB}=0$, ${\Phi}_i=0$). This solution is clearly given by the generators $L_{AB}$ of the irreducible representation $(\frac{L}{2},\frac{L}{2})$  of $SO(4)$, viz
\begin{eqnarray}
D_{AB}=L_{AB}.\label{sol}
\end{eqnarray}
As it turns out this is also the absolute minimum of the model. By expanding $D_{AB}$ around this vacuum as $D_{AB}=L_{AB}+A_{AB}$ and substituting back into the action (\ref{model}) we obtain a $U(1)$ gauge field $A_{AB}$ on ${\bf S}^2_L{\times}{\bf S}^2_L$ with the correct transformation law under the action of the group, namely $A_{AB}{\longrightarrow}UA_{AB}U^{-1}+U{\cal L}_{AB}U^{-1}$. The matrices $D_{AB}$ are thus  the covariant derivatives on ${\bf S}^2_L{\times}{\bf S}^2_L$. The curvature $F_{AB,CD}$ in terms of $A_{AB}$ takes the usual form $F_{AB,CD}=i{\cal L}_{AB}A_{CD}-i{\cal L}_{CD}A_{AB}+{f}_{ABCDEF}A_{EF}+i[A_{AB},A_{CD}]$. The normal scalar fields in terms of $A_{AB}$ are on the other hand given by $8\sqrt{c_2}({\Phi}_1+{\Phi}_2)=L_{AB}A_{AB}+A_{AB}L_{AB}+A_{AB}^2$ and $16\sqrt{c_2}({\Phi}_1-{\Phi}_2)={\epsilon}_{ABCD}(L_{AB}A_{CD}+A_{AB}L_{CD}+A_{AB}A_{CD})$. 

We can verify this conclusion explicitly by  introducing the  matrices $D^{(1)}_{a}=L_a^{(1)}+A_a^{(1)}$ and $ D^{(2)}_{a}=L_a^{(2)}+A_a^{(2)}$ defined by 
\begin{eqnarray}
D^{(1)}_{a}{\equiv}-\frac{1}{2}\left[\frac{1}{2}\epsilon_{abc}D_{bc}+D_{a4}\right]~,~D^{(2)}_{a}{\equiv}-\frac{1}{2}\left[\frac{1}{2}\epsilon_{abc}D_{bc}-D_{a4}\right].\label{la} 
\end{eqnarray}
Clearly $D_a^{(1)}$ ($A_a^{(1)}$)  and $D_a^{(2)}$ ($A_a^{(2)}$) are the components of $D_{AB}$ ($A_{AB}$) on the two spheres respectively. The curvature becomes $F^{(i,j)}_{ab}=i\mathcal{L}^{(i)}_{a}A^{(j)}_{b}-i\mathcal{L}^{(j)}_{b}A^{(i)}_{a}+{\delta}_{ij}\epsilon_{abc}A^{(i)}_{c}+i[A_a^{(i)},A_b^{(j)}]$ whereas the normal scalar fields become $2\sqrt{c_2}{\Phi}_i=(D_a^{(i)})^2-c_2=L_a^{(i)}A_a^{(i)}+A_a^{(i)}L_a^{(i)}+(A_a^{(i)})^2$. In terms of this three dimensional notation the action (\ref{model}) reads
\begin{eqnarray}
&&S=S^{(1)}+S^{(2)}+\frac{1}{2g^2}Tr_{L}\left(F_{ab}^{(1,2)}\right)^2.\label{ac}
\end{eqnarray}
$S^{(1)}$ and $S^{(2)}$ are the actions for the $U(1)$ gauge fields $A_a^{(1)}$ and $A_a^{(2)}$ on a single fuzzy sphere ${\bf S}^2_L$. They are given by
\begin{eqnarray}
S^{(i)}
&=&\frac{1}{4g^2}Tr_{L}\left(F_{ab}^{(i,i)}\right)^2-\frac{1}{2g^2}{\epsilon}_{abc}Tr_L\left[\frac{1}{2}F_{ab}^{(i,i)}A_c^{(i)}-\frac{i}{6}[A_a^{(i)},A_b^{(i)}]A_c^{(i)}\right]+\frac{2m^2}{g^2}Tr_{L}{\Phi}_i^2.\nonumber\\&&
\end{eqnarray}
It is immediately clear that in the continuum limit  $L{\longrightarrow}{\infty}$ the action (\ref{ac}) describes the interaction of a genuine $4-$d gauge field with the normal scalar fields ${\Phi}_i=n_a^{(i)}A_a^{(i)}$ where $n_a^{(i)}$ is the unit normal vector to the $i$-th sphere. The parameter $m$ is precisely the mass of these scalar fields. Let us also remark that in this limit the $3-$dimensional fields $A_a^{(i)}$ decompose as $A_a^{(i)}=(A_a^{(i)})^T+n_a^{(i)}{\Phi}_i$ where $(A_a^{(i)})^T$ are the tangent $2-$dimensional gauge fields. Since the differential calculus on ${\bf S}_L^2 \times {\bf S}_L^2$ is intrinsically $6-$dimensional we can not decompose the fuzzy gauge field in a similar (gauge-covariant) fashion and as a consequence we can not write an action on the fuzzy ${\bf S}_L^2 \times {\bf S}_L^2$ which will only involve the desired $4-$dimensional gauge field.

\section{Quantum effective potential}
The partition function of the theory depends  on $3$
parameters, the Yang-Mills coupling constant $g$, the mass $m$ of the normal scalar fields, and the size $L$ of the matrices, viz
\begin{eqnarray}
Z_{L}\left[J,g,m\right]=\int \prod_{A<B=1}^4\left[dX_{AB}\right]e^{-S[X]+Tr_L\left[J_{AB}X_{AB}\right]}.
\end{eqnarray}
In the background field method the field is decomposed as $X_{AB}=D_{AB}+Q_{AB}$ where $D_{AB}$ is the background we are interested in studying and $Q_{AB}$ stands for the fluctuation field. We add the usual gauge fixing and Faddeev-Popov terms given by
\begin{eqnarray}
 S_{g.f}+S_{gh} = 
-\frac{1}{32g^2}Tr_L \frac{[D_{AB},Q_{AB}]^2}{\xi}+\frac{1}{16g^2}Tr_L  c[D_{AB},[D_{AB},b]].
\label{ghost}
\end{eqnarray}
 Performing the Gaussian path integral we obtain the one-loop effective action
\begin{equation}
\Gamma\left[D_{AB}\right]=S\left[D_{AB}\right]+\frac{1}{2}Tr_{6}TR\log{\Omega}_{ABCD}-TR\log\mathcal{D}_{AB}^{2}.
\label{effaction}
\end{equation}
${\Omega}_{ABCD}$ is defined by
\begin{eqnarray}
{\Omega}_{ABCD}&=&\frac{1}{2}\mathcal{D}_{EF}^2{\delta}_{AB,CD}-\left(1-\frac{1}{\xi}\right)\mathcal{D}_{AB}\mathcal{D}_{CD}-2i\mathcal{F}_{ABCD}+\frac{4 m^2}{L^{2}_{AB}}\Omega^{\left(1\right)}_{ABCD},
\end{eqnarray}
where $\delta_{AB,CD}=\delta_{AC}\delta_{BD}-\delta_{AD}\delta_{BC}$, and 
\begin{eqnarray}
\Omega^{\left(1\right)}_{ABCD}&=&(D^2_{EF}-L^2_{EF})\delta_{AB,CD}+\frac{1}{2}(\epsilon_{EFGH}D_{EF}D_{GH})\epsilon_{ABCD}\nonumber\\
&&-\mathcal{D}_{AB}\mathcal{D}_{CD}
-\widetilde{\mathcal{D}}_{AB}\widetilde{\mathcal{D}}_{CD}
+4D_{AB}D_{CD}+4\widetilde{D}_{AB}\widetilde{D}_{CD}.
\end{eqnarray}
The notation ${\mathcal D}_{AB}$ and ${\mathcal F}_{ABCD}$ means that
the covariant derivative $D_{AB}$ and the curvature $F_{ABCD}$ act by
commutators, i.e ${\cal D}_{AB}(M)=[D_{AB},M]$, ${\cal
F}_{ABCD}(M)=[F_{ABCD},M]$ where $M$ is an element of
$Mat_{(L+1)^{2}}$. Wwe have also introduced the notation $\widetilde{D}_{AB}\equiv\frac{1}{2}\epsilon_{ABCD}D_{CD}$. $TR$ is the trace over the $4$ indices corresponding to the left and  right actions of  operators on matrices. $Tr_{6}$ is the trace associated with the action of $SU(2){\times}SU(2)$.

The main goal of this article is to check the stability of the
solution (\ref{sol}), in other words to check whether or not the fuzzy space
${\bf S}^2_L{\times}{\bf S}^2_L$ is stable under quantum fluctuations. Towards this end it is sufficient to consider only the  background field $D_{AB}=\phi L_{AB}$ where the order parameter $\phi$ plays the role of the radius of the two spheres of ${\bf S}^2_L{\times}{\bf S}^2_L$. Therefore the computation of the effective action reduces to the computation of the effective potential $V_{\rm eff}(\phi)\equiv \Gamma [ \phi L_{AB} ]$. The classical potential is given by

\begin{equation}
V{\equiv}S[{\phi}L_{AB}]=\frac{L(L+2)}{g^2}\left(\frac{1}{4}\phi^{4}-\frac{1}{3}\phi^{3}+\frac{1}{4}m^{2}\left(\phi^{2}-1\right)^{2}\right).
\end{equation}  
The effective potential (in the gauge $\xi=1$) is given by 

\begin{eqnarray}
V_{\rm eff}&=&V+\frac{1}{2}Tr_6TR\log{\phi}^2-TR\log{\phi}^2+\frac{1}{2}Tr_6TR\log\tilde{\Omega}_{ABCD}\nonumber\\
&=&V+4(L+1)^4\log\phi +\frac{1}{2}Tr_6TR\log\tilde{\Omega}_{ABCD}.\label{la1}
\end{eqnarray}
We are only interested in the ${\phi}-$dependence of the operator $\tilde{\Omega}$ which is defined by
\begin{eqnarray}
\tilde{\Omega}_{ABCD}=\frac{1}{2}{\cal L}_{EF}^2{\delta}_{AB,CD}+2i\left(1-\frac{1}{\phi}\right)f_{ABCDEF}{\cal L}_{EF}+\frac{4m^2}{L_{AB}^2}\tilde{\Omega}_{ABCD}^{(1)},
\end{eqnarray}  
where
\begin{eqnarray}
\tilde{\Omega}^{\left(1\right)}_{ABCD}&=&\left(1-\frac{1}{{\phi}^2}\right)L_{EF}^2\delta_{AB,CD}-\mathcal{L}_{AB}\mathcal{L}_{CD}-\widetilde{\mathcal{L}}_{AB}\widetilde{\mathcal{L}}_{CD}+4L_{AB}L_{CD}+4\widetilde{L}_{AB}\widetilde{L}_{CD}.\nonumber \\&&
\end{eqnarray}
We will need to use the following identities
\begin{eqnarray}
X_{AB}Y_{AB}&=&4\left(X^{\left(1\right)}_{a}Y^{\left(1\right)}_{a}+X^{\left(2\right)}_{a}Y^{\left(2\right)}_{a}\right),\nonumber\\
f_{ABCDEF}Tr\left[X_{AB}Y_{CD}Z_{EF}\right]&=&16\epsilon_{abc}Tr\left[X^{\left(1\right)}_{a}Y^{\left(1\right)}_{b}Z^{\left(1\right)}_{c}+X^{\left(2\right)}_{a}Y^{\left(2\right)}_{b}Z^{\left(2\right)}_{c}\right].
\end{eqnarray}
The matrices $X^{\left(i\right)}_{a}$ ($Y^{\left(i\right)}_{a}$) are related to the matrices $X_{AB}$ ($Y_{AB}$) by equations of the form (\ref{la}). Using these identities we can express the last term in (\ref{la1}) in the following way

\begin{eqnarray}
\frac{1}{2}Tr_6TR\log\tilde{\Omega}_{ABCD}&=&\int dX_{AB} e^{-Tr X_{AB}\tilde{\Omega}_{ABCD}X_{CD}}\nonumber\\
&=&\left[\int
dX_a^{(1)}~e^{-2Tr X_a^{(1)}\tilde{\Omega}_{ab}X_b^{(1)}}\right]^{2}\nonumber\\
&=&Tr_3TR\log\tilde{\Omega}_{ab}.\label{4.11}
\end{eqnarray}
The contributions coming from the two spheres are equal and hence the factor of $1$ (instead of $\frac{1}{2}$) in front of the last logarithm. $Tr_3$ is the trace associated with the action of $SU(2)$ on the two dimensional sphere. The Laplacian $\tilde{\Omega}_{ab}$ is defined by
\begin{eqnarray}
\tilde{\Omega}_{ab}&=&2\mathcal{L}^{2}_{AB}\delta_{ab}+16\left(1-\frac{1}{\phi}\right)i{\epsilon}_{abc}{\cal L}_c^{(1)}+8m^2\tilde{\Omega}^{(1)}_{ab},\nonumber\\
\tilde{\Omega}^{(1)}_{ab}&=&4P_{ab}^{(1)}-\frac{1}{c_2}{\cal L}_a^{(1)}{\cal L}_b^{(1)}+2\left(1-\frac{1}{{\phi}^2}\right){\delta}_{ab}.
\end{eqnarray}
$P_{ab}^{(1)}$ is the normal projector on the fuzzy sphere defined by
${P}_{ab}^{(1)}=x^{\left(1\right)}_{a}x^{\left(1\right)}_{b}$ where
$x^{\left(1\right)}_{a}$ are the coordinate operators defined in
(\ref{cor}) with $R_1=R_2=1$. The presence of this projector means in
particular that we can not diagonalize in the polarization tensors
basis. However, in order to have an idea of the phase structure of the model, we can expand around $m=0$. This approximation was more than sufficient in the two-dimensional case as discussed in great detail in \cite{PRY}. Therefore it is convenient to separate the logarithm term as
\begin{equation}
\log\tilde{\Omega}^{}_{ab}=\log\tilde{\Omega}^{(0)}_{ab}+\log\left(1+8m^2\left(\frac{1}{\tilde{\Omega}^{(0)}}\right)_{ac}\tilde{\Omega}^{(1)}_{cb}\right).
\end{equation}
$\tilde{\Omega}^{(0)}_{ab}$ is clearly equal to $\tilde{\Omega}_{ab}$ when $m^2=0$. This operator can be trivially diagonalized in the vector polarization tensors basis $(\hat{Y}^{j_1M_1}_{l_1})_{a}$ on the first sphere tensor product the scalar polarization tensors basis $\hat{Y}_{l_2m_2}$ on the second sphere.  Indeed by introducing the total angular momentum on the two-dimensional sphere ${\cal J}_a^{(1)}={\cal L}_a^{(1)}+{\theta}_a^{(1)}$ where ${\theta}_a^{(1)}$ are the generators of $SU(2)$ in the spin $1$ irreducible representation we can rewrite $\tilde{\Omega}^{(0)}_{ab}$ in the following form
\begin{equation}\label{omega}
\frac{1}{8}\tilde{\Omega}^{(0)}_{ab}=({\cal L}_c^{(1)})^2 {\delta}_{ab}+({\cal L}_c^{(2)})^2 {\delta}_{ab}-\left(1-\frac{1}{\phi}\right)\big[({\cal J}_c^{(1)})^2_{ab}-({\cal L}_c^{(1)})^2 {\delta}_{ab}-2{\delta}_{ab}\big].
\end{equation}
Hence it is convenient to use the following expansion for the matrices $X^{\left(1\right)}_{a}$ in (\ref{4.11})
\begin{equation}
X^{\left(1\right)}_{a}=\sum_{j_{1}M_{1}\ell_{1}}\sum_{\ell_{2}m_{2}}q^{j_{1}M_{1}\ell_{1}}_{\ell_{2}m_{2}}\left(\hat{Y}^{j_{1}M_{1}}_{\ell_{1}}\right)_{a}\otimes\hat{Y}_{\ell_{2}m_{2}}.
\end{equation}
Thus
\begin{eqnarray}
Tr_{3}TR
\log\tilde{\Omega}^{(0)}_{ab}=\sum_{\ell_{1}j_{1}\ell_{2}}\left(2j_{1}+1\right)\left(2\ell_{2}+1\right)\log\left[1-2\left(1-\frac{1}{\phi}\right)\frac{j_{1}\left(j_{1}+1\right)-\ell_{1}\left(\ell_{1}+1\right)-2}{\ell_{1}\left(\ell_{1}+1\right)+\ell_{2}\left(\ell_{2}+1\right)}\right]\nonumber \\&&
\label{contri1}
\end{eqnarray}
In the limit $L{\longrightarrow}{\infty}$ it is easily verifiable (for example by making an expansion in $1-\frac{1}{\phi}$) that this term is subleading compared to $L^4$. The second contribution in the limit $m{\longrightarrow}0$ is given by
\begin{eqnarray}\label{massdelta}
&&Tr_3TR\log\left(1+8m^2\left(\frac{1}{\tilde{\Omega}^{(0)}}\right)_{ac}\tilde{\Omega}^{(1)}_{cb}\right) \approx  32m^2Tr_3TR\left(\frac{1}{\tilde{\Omega}^{(0)}}\right)_{ac}x_c^{(1)}x_b^{(1)}\nonumber\\&&\qquad\qquad\qquad-\frac{8m^2}{c_2}Tr_3TR\left(\frac{1}{\tilde{\Omega}^{(0)}}\right)_{ac}{\cal L}_c^{(1)}{\cal L}_b^{(1)}
+16m^2\left(1-\frac{1}{{\phi}^2}\right)Tr_3TR\;\left(\frac{1}{\tilde{\Omega}^{(0)}}\right)_{ab}.\nonumber\\&&\label{contri2}
\end{eqnarray}
In the large $L$ limit it is possible to show (see the appendix) that  all terms in (\ref{contri2}) are subleading compared to the $L^4$ behaviour seen in the second term in (\ref{la1})  and hence the full one-loop quantum contribution to the effective potential is given by the logarithmic potential in (\ref{la1}). Thus as long as we are in the region of the phase space near  $m\approx 0$ the effective potential behaves in the large $L$ limit as follows
\begin{equation}
\frac{V_{\rm eff}}{4L^4}=\frac{1}{4g^2L^2}\left(\frac{1}{4}\phi^{4}-\frac{1}{3}\phi^{3}+\frac{1}{4}m^{2}\left(\phi^{2}-1\right)^{2}\right)+\log\phi.\label{4.8}
\end{equation}
This result is to be compared with the quantum effective potential for $U(1)$ gauge fields on a single fuzzy sphere ${\bf S}^2_L$ computed in \cite{PRY} which is given explicitly by
 \begin{eqnarray}
\frac{V_{\rm eff}}{L^2}=\frac{1}{2g^2} \bigg[\frac{1}{4}{\phi}^4-\frac{1}{3}{\phi}^3+\frac{1}{4}m^2({\phi}^2-1)^2\bigg]+\log{\phi}.
\end{eqnarray}
\begin{eqnarray}
V_{\rm eff}(\phi)=2c_2N^2{\alpha}^4 \bigg[\frac{1}{4}{\phi}^4-\frac{1}{3}{\phi}^3\bigg]+4c_2\log{\phi}+{\rm subleading~terms}.
\end{eqnarray}

\section{The ${\bf S}^2_L{\times}{\bf S}^2_L$-to-matrix phase transition}
The second term in the potential (\ref{4.8}) is not convex. This
implies that there is a competition between the classical potential
and the logarithmic term which  depends on the values of $m$ and
$g$. The equation of motion $\frac{\partial{V_{\rm
      eff}}}{\partial{\phi}}=0$ will admit in general two real
solutions where the one with the least energy can be identified with
the fuzzy ${\bf S}^2_L{\times}{\bf S}^2_L$ solution
(\ref{sol}). This equation of motion reads
\begin{eqnarray}
(1+m^2){\phi}^4-{\phi}^3-m^2{\phi}^2+4g^2L^2=0.
\end{eqnarray}
The quantum solution is found to be very close to $1$, viz
\begin{eqnarray}
\phi=1-\frac{4g^2L^2}{1+2m^2}+O((g^2L^2)^2).
\end{eqnarray}
However this is only true up to an upper value of the gauge coupling constant $g$ (for every fixed value of $m$) beyond which the equation of motion ceases to have any real solutions. At this value the fuzzy ${\bf S}^2_L{\times}{\bf S}^2_L$ collapses under the effect of quantum fluctuations and we cross   to a pure matrix phase. In other words we can not define a gauge theory everywhere in the phase space. As we will see below when the mass $m$ is sent to infinity it is more difficult to reach the matrix phase and hence the presence of the mass makes the fuzzy ${\bf S}^{2}_L\times {\bf S}^{2}_L$ solution (\ref{sol}) more stable. 

The critical value can be computed by requiring that both the first and the second derivatives of the potential $V_{\rm eff}$ with respect to $\phi$  vanish. In other words, for every fixed value of $m$ the critical point is defined at the point $(g_{*},m)$ of the phase space where we go from a bounded potential to an unbounded potential. Solving for the critical value we get the results
\begin{eqnarray}
{\phi}_{*}=\frac{3}{8(1+m^2)}\bigg[1+\sqrt{1+\frac{32m^2(1+m^2)}{9}}\bigg],\label{1}
\end{eqnarray}
and
\begin{eqnarray}
2g^2_{*}L^2=-\frac{1}{2}(1+m^2){\phi}_{*}^4+\frac{1}{2}{\phi}_{*}^3+\frac{m^2}{2}{\phi}_{*}^2.\label{2}
\end{eqnarray}
In the particular case of $m^2=0$ the critical value is
\begin{equation}
g^2_{*}L^2=\frac{1}{2}\left(\frac{3}{8}\right)^3.
\end{equation}
Extrapolating to large values of the  mass  ($m{\longrightarrow}{\infty}$) we obtain the scaling behaviour 
\begin{eqnarray}
g^2_{*}L^2=\frac{m^2+\sqrt{2}-1}{16}. \label{pre1}
\end{eqnarray}
In figure $1$ we plot the phase diagram defined by this
equation\footnote{Notice that if we allow $m^2$ to take negative values, the gauge coupling  constant $g_{*}^2$ will be a more complicated function of $m^2$. However we are only interested in positive values of $m^2$ for which the behaviour of $g_{*}^2$ as a function of $m^2$ is the straight line (\ref{pre1}) which can be deduced from the large $m^2$ behaviour of (\ref{1}) and (\ref{2}). }. As
we increase the value of the coupling constant $g$ (for a fixed  value
of $m^2$) there exists a {\it critical point} $g_{*}$ where the fuzzy
${\bf S}_L^{2}\times {\bf S}_L^{2}$ solution becomes unstable and thus
the minimum (\ref{sol}) disappears. Similarly as the value of the
mass squared $m^2$ increases (for a fixed  value of the coupling
constant $g$) there is a {\it critical point} $m^2_{*}$ where ${\bf
  S}_L^{2}\times {\bf S}_L^{2}$ collapses. Clearly the value of
$m^2_{*}$ is found by inverting equation (\ref{pre1}) , viz
\begin{eqnarray}
m^2_{*}=16g^2L^2+1-\sqrt{2}. 
\end{eqnarray}
Finally we remark that as the value of
$m^2$ increases it is more difficult to reach the transition point, in
fact when $m^2{\longrightarrow}{\infty}$ the critical value $g^2_{*}$
approaches infinity.

\begin{figure}[h]
\begin{center}
\includegraphics[width=7cm,angle=-90]{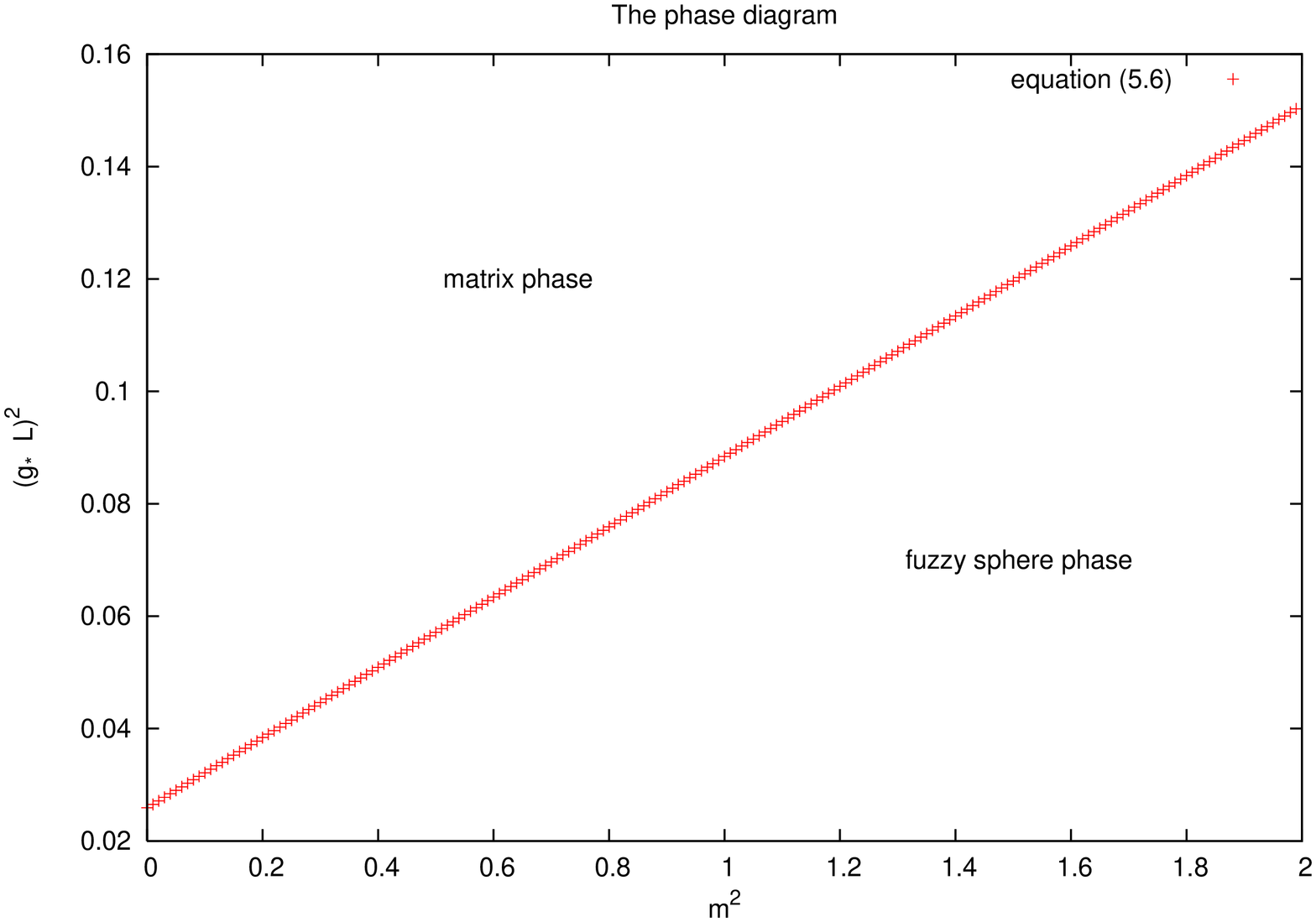} 
\caption{{ The ${\bf S}^2_L{\times}{\bf S}^2_L$-to-matrix critical line.}}
\end{center}
\end{figure}

\section{Conclusion}

We have described the qualitative behaviour of a first order phase
transition which occurs in a $U(1)$ gauge theory on ${\bf
  S}^{2}_{L}\times {\bf S}^{2}_{L}$. Using the one-loop effective
potential (\ref{4.8}) of this theory we found that there exists values
of the gauge coupling constant $g$ and the mass $m$ for which the
fuzzy ${\bf S}^{2}_{L}\times {\bf S}^{2}_{L}$ solution (\ref{sol}) is
not stable. Thus for these values a $U(1)$ gauge theory on ${\bf
  S}^{2}_L\times {\bf S}^{2}_L$ is not well defined. This means in
particular that the model (\ref{model}) can be used to approximate
$U(1)$ gauge field theories on ${\bf S}^2\times {\bf S}^2$ only deep
inside the fuzzy sphere phase. However it is obvious from the critical
line (\ref{pre1}) that when the mass  $m$ of the two normal scalar
fields on  ${\bf S}^{2}_{L}\times {\bf S}^{2}_{L}$ goes to infinity it
is more difficult to reach the transition line. Therefore we can say
that our main goal of defining a nonperturbative  regularization of a
$U(1)$ gauge theory on ${\bf S}^{2} \times {\bf S}^{2}$ is
achieved. Generalization to $U(n)$ with and without fermions should be
straightforward as long as we are only interested in the effective potential.

\paragraph{Acknowledgements}
The authors P. Castro-Villarreal and R. Delgadillo-Blando  would like
to thank Denjoe O'Connor for his supervision through the course of
this study. B. Ydri would like to thank  Denjoe O'Connor for his
extensive discussions and critical comments while this research was in
progress. The work of P. C. V. was supported by
CONACYT M\'exico Grant No.43891-Y. The work of R. D. B. is supported by CONACYT M\'exico.

\appendix
\section{Evaluation of $I_{1}$, $I_{2}$ and $I_{3}$.}
In this appendix we show that  the $3$ terms in (\ref{massdelta}) are subleading compared to $L^4$. Let us define
\begin{eqnarray}
&&I_{1}=\frac{1}{L^{4}}Tr_{3}TR\left(\frac{1}{\tilde{\Omega}^{(0)}}\right)_{ab},~~~
I_{2}=\frac{1}{L^{6}}Tr_{3}TR\left(\frac{1}{\tilde{\Omega}^{(0)}}\right)_{ab}\mathcal{L}^{\left(1\right)}_{b}\mathcal{L}^{\left(1\right)}_{c},~~~\nonumber\\
&&\qquad\quad\quad\quad I_{3}=\frac{1}{L^{4}}Tr_{3}TR\left(\frac{4}{\tilde{\Omega}^{(0)}}\right)_{ab}x^{\left(1\right)}_{b}x^{\left(1\right)}_{c}.
\end{eqnarray}
We evaluate these traces by using the base of polarization tensors 
$\left(\hat{Y}^{j_{1}M_{1}}_{\ell_{1}}\right)_{a}\otimes\hat{Y}_{\ell_{2}m_{2}}$.
Using the identity
$\mathcal{L}_a\mathcal{L}_b=\mathcal{L}^2\delta_{ab}-(\theta\cdot\mathcal{L})_{ab}-(\theta\cdot\mathcal{L})^2_{ab}$
the eigenvalues of the operator $\mathcal{L}_a\mathcal{L}_b$ are given by
\begin{eqnarray}
\eta_{\ell_{1}
  j_{1}}&=&\frac{1}{4}\left(j_{1}\left(j_{1}+1\right)-\ell_{1}\left(\ell_{1}+1\right)\right)^{2}-\frac{1}{2}\left(j_{1}\left(j_{1}+1\right)+\ell_{1}\left(\ell_{1}+1\right)\right),
\end{eqnarray}
whereas the eigenvalues of $\tilde{\Omega}^{(0)}_{ab}$ given by
(\ref{omega}) are 
\begin{eqnarray}
\lambda^{\ell_{1}j_{1}}_{\ell_{2}}&=&8\left(\ell_{1}\left(\ell_{1}+1\right)+\ell_{2}\left(\ell_{2}+1\right)\right)+8\frac{1-\phi}{\phi}\left(j_{1}\left(j_{1}+1\right)-\ell_{1}\left(\ell_{1}+1\right)-2\right).
\end{eqnarray}
Using these facts the two quantities $I_{1}$ and $I_{2}$ can be shown to be  given by 
\begin{equation}\label{I1yI2}
I_{1}=\frac{1}{L^{4}}\sum_{\ell_{1}j_{1}\ell_{2}}\frac{\left(2l_{2}+1\right)\left(2j_{1}+1\right)}{\lambda^{\ell_{1}j_{1}}_{\ell_{2}}},~~~~~I_{2}=-\frac{1}{L^{6}}\sum_{\ell_{1}j_{1}\ell_{2}}\frac{\left(2\ell_{2}+1\right)\left(2j_{1}+1\right)\eta_{\ell_{1} j_{1}}}{\lambda^{\ell_{1}j_{1}}_{\ell_{2}}}.
\end{equation}
 In order to evaluate $I_3$ we notice the fact that $x^{(i)}_a$ is proportional to
$\left(\hat{\bf Y}^{00}_{1}\right)_a$ thus by using the algebra of vectorial
polarization tensors we  get the  identity
\begin{eqnarray}
&&\mbox{Tr}\left\{\left({\bf Y}^{jM}_{\ell}\cdot {\bf
  Y}^{00}_1\right)\left({\bf Y}^{00}_1\cdot {\bf
  Y}^{+\;jM}_{\ell}\right)\right\}=(L+1)(2\ell+1)
\left\{\begin{array}{ccc}
        1&\ell&j \\
	\frac{L}{2}&\frac{L}{2}&\frac{L}{2}
       \end{array}\right\}^2.
\end{eqnarray}
The final result for $I_{3}$ is 
\begin{equation}\label{I3}
I_{3}=\frac{2}{L^{4}}\sum_{\ell_{1}j_{1}\ell_{2}}\frac{2l_{2}+1}{\lambda^{\ell_{1}j_{1}}_{\ell_{2}}}\left[\left(L+1\right)\left(2\ell_{1}+1\right)\left\{\begin{array}{ccc}
1 & \ell_{1} & j \\
\frac{L}{2} &\frac{L}{2} &\frac{L}{2} 
\end{array}
\right\}^{2}\right].
\end{equation}
The large $L$ behaviour of $I_1$, $I_2$ and $I_3$ can be studied with the help of the different identities of \cite{VKM}. The first sum in
(\ref{I1yI2}) diverges at most as
$L^2$ in
the continuum  $L\to\infty$ limit  and hence $I_1$ converges to zero as $1/L^2$. On the other hand the sum in $I_{2} $ behaves at
most as $L^{4} $ thus the whole expression goes to zero as $1/L^2$. For
$I_3$ we can check that the sum goes as $L$, i.e $I_3$ appraoches $0$ as $1/L^3$.

\bibliographystyle{unsrt}

\begin{thebibliography}{99}



\bibitem{PRY} 
P. Castro-Villarreal, R. Delgadillo-Blando, Badis Ydri. {\it `` A
  gauge invariant UV-IR mixing and the corresponding phase transition
  for $U\left(1\right)$ fields on fuzzy sphere''}, Nucl.Phys.B {\bf
  704} (2005) 111-153, [hep-th/0405201].

\bibitem{nishimura} T.Azuma,S.Bal,K.Nagao,J.Nishimura, JHEP {\bf 05} (2004) 005, [hep-th/0401038].

\bibitem{steinacker}
C.Klim\v cik, Commun.Math.Phys. {\bf 199} (1998) 257, [hep-th/9710153];\\
H.Grosse,P.Pre\v snajder, Lett.Math.Phys. {\bf 46} (1998) 61-69; H.Grosse,P.Pre\v snajder, [hep-th/9805085];\\
U.C.Watamura,S.Watamura, Commun.Math.Phys. {\bf 212} (2000) 395, [hep-th/9801195]; \\
S.Iso,Y.Kimura,K.Tanaka,K.Wakatsuki, Nucl.Phys.B {\bf 604} (2001) 121-147, [hep-th/0101102];\\
T.Imai,Y.Kitazawa,Y.Takayama,D.Tomino, Nucl.Phys.B {\bf 665} (2003) 520, [hep-th/0303120]; \\
H.Steinacker, Nucl.Phys.B {\bf 679} (2004) 66, [hep-th/0307075];\\
D.O'Connor, Mod.Phys.Lett.{\bf A} 18 (2003) 2423-2430.


\bibitem{ydri}
Monte Carlo Simulation of NC Gauge Field on The Fuzzy Sphere , in progress.

\bibitem{madore} J.Madore, Class.Quant.Grav.{\bf 9} (1992) 69-88; \\
J.Hoppe,MIT PhD thesis (1982); J.Hoppe,S.T.Yau, Commun.Math.Phys. {\bf 195} (1998) 67-77.

\bibitem{peter}H.Grosse,C.Klim\v cik,P.Pre\v snajder,
  Int.J.Theor.Phys. {\bf 35}, (1996) 231, [hep-th/9505175].

\bibitem{thesis}B.Ydri, Fuzzy Physics, PhD Thesis, hep-th/0110006. 

\bibitem{scalar} S.Vaidya, Phys.Lett.B {\bf 512} (2001) 403-411, [hep-th/0102212];\\
C.S.Chu,J.Madore,H.Steinacker, JHEP {\bf 08} (2001) 038, [hep-th/0106205]; \\
B.P.Dolan,D.O'Connor,P.Presnajder, JHEP {\bf 03} (2002) 013, [hep-th/0109084];\\ 
H.Steinacker, JHEP {\bf 03} (2005) 075, [hep-th/0501174].

\bibitem{xavier} X.Martin, JHEP {\bf 04} (2004) 077, [hep-th/0402230]; X.Martin, Mod.Phys.Lett.A {\bf 18} (2003) 2389-2396.

\bibitem{bieten}W.Bietenholz,F.Hofheinz and J.Nishimura, JHEP {\bf 06} (2004) 042 [ hep-th/0404020].


\bibitem{scalar4d} S.Vaidya,B.Ydri, Nucl.Phys.B {\bf 671} (2003) 401-431, [hep-th/0305201];\\
S.Vaidya,B.Ydri, [hep-th/0209131];\\
 H.Grosse,A.Strohmaier, Lett.Math.Phys. {\bf 48} (1999) 163-179, [hep-th/9902138];
 G.Alexanian,A.P.Balachandran,G.Immirzi,B. Ydri, J.Geom.Phys. {\bf 42} (2002) 28-53, [hep-th/0103023];\\
J.Medina,D.O'Connor, JHEP {\bf 11} (2003) 051, [hep-th/0212170].


\bibitem{othergaugein4d} 
Y.Kimura, Nucl.Phys.B {\bf 637} (2002) 177-198, [hep-th/0204256];\\
T.Imai,Y.Kitazawa,Y.Takayama,D.Tomino, Nucl.Phys.B {\bf 679} (2004) 143-167, [hep-th/0307007];\\
T.Imai,Y.Takayama, Nucl.Phys.B {\bf 686} (2004) 248-260, [hep-th/0312241];\\
T.Azuma,S.Bal,K.Nagao,J.Nishimura, [hep-th/0405277];\\
H.Grosse,H.Steinacker, Nucl.Phys.B {\bf 707} (2005) 145-198, [hep-th/0407089];\\
W.Behr,F.Meyer,H.Steinacker, [hep-th/0503041];\\



\bibitem{connes}
A.Connes, {\it Noncommutative Geometry}, Academic Press,
London,1994. Landi, {\it An introduction to noncommutative spaces and their geometry}, springer (1997).
J.M. Gracia-Bondia, J. C. Varilly, H. Figueroa, {\sl Elements of Noncommutative Geometry}, Birkhauser (2000).
J.Madore,{\it An Introduction to Noncommutative Differential Geometry and its Physical Applications }, Cambridge Press (1995).

\bibitem{VKM}
D. A. Varshalovich, A. N. Moskalev, V. K. Khersonky, 
{\sl Quantum Theory of Angular Momentum: Irreducible Tensors, Spherical Harmonics, Vector Coupling Coefficients, 3nj Symbols}, Singapore, Singapore. World Scientific (1998).

\bibitem{badis}
B.Ydri, Nucl.Phys. B {\bf 690} (2004) 230-248, [hep-th/0403233];
B. Ydri, Phys.Letter A {\bf 19} (2004) 2205, [hep-th/0405208].


\end{thebibliography}

\end{document}